\documentclass[11pt]{article}

\usepackage{amsmath,amssymb,amscd}
\usepackage{cite}
\usepackage{amsmath, amsthm, amssymb, amsfonts}
\usepackage{dsfont}
\usepackage{float}
\usepackage[a4paper]{geometry}
\usepackage{color}
\usepackage{booktabs}
\usepackage{lineno}
\usepackage{graphicx}
\usepackage[mathscr]{eucal}

\large\normalsize

\title{Evolutionary rationality of risk preference}
\author{Song-jia Fan$^{1,2}$,Yi Tao$^{1,2}$ and Cong Li$^{3}$\footnote{Author for correspondence, and e-mail:
		congli@nwpu.edu.cn}
	\\ $^1$Key Laboratory of Animal Ecology and Conservation Biology,\\
	Centre for Computational and Evolutionary Biology, \\
	Institute of Zoology, Chinese Academy of Sciences, \\
	Beijing, P.R. China \\
	$^2$University of Chinese Academy of Sciences, \\ Beijing, P.R. China \\
	$^3$School of Ecology and Environment, Northwestern Polytechnical University, \\ Xi'an , P.R. China}

\date{}

\begin{document}
	
	\maketitle
	
	\newpage
	
	\section*{Significance}
	Risk does not discriminate. It tests all individuals with the
	uncertain future consequences of present actions, regardless of
	species or intelligence. And selection takes place in the process
	ever since ancient time. The kinds of risk attitude that survived
	evolution must be rational in the sense that they adapt to different
	intensity of selection in order to maximize the winning chance of their
	carriers. Here we present the theory of evolutionary rationality,
	proposing that for the human society, the mental equivalent of
	the intensity of selection, the attention degree that is given to a risky
	event, determines our attitude towards it. An evolutionary rational
	person will adopt extreme risk aversion for crucial decisions, and
	swing towards risk neutral for trivial decisions.
	
	\section*{Abstract}
	Selection shapes all kinds of behaviors, including how we make
	decisions under uncertainty. The risk attitude reflected from it
	should be simple, flexible, yet consistent. In this paper we engaged
	evolutionary dynamics to find the decision making rule concerning
	risk that is evolutionarily superior, and developed the theory of
	evolutionary rationality. We highlight the importance of selection
	intensity and fitness, as well as their equivalents in the human
	mind, named as attention degree and meta-fitness, in the decision
	making process. Evolutionary rationality targets the maximization of the geometric mean of meta-fitness (or fitness), and attention
	degree (or intensity of selection) holds the key in the change of
	attitude of the same individual towards different events and under
	varied situations. Then as an example, the Allais paradox is presented to show the application of evolutionary rationality, in which the anomalous choices made by the majority of people can be well justified by a simple change in the attention degree.

	\newpage
	
	\section{Introduction}
	
	How much would you pay for a bet with an expected return of \$1
	million? A lot, I guess? However, St. Petersburg Paradox
	\cite{bernoulli1954exposition} shows that even if the expected
	return of a bet is as high as infinity, when the probability of
	winning big prize is low, most people will not spend big money on
	it. The paradox reflects the existence of risk preference, which is
	ubiquitous in both the human and the animal world. We all have an
	attitude towards risk, but our risk preference is not immutable.
	Financial speculators buy insurance for their cars and houses
	\cite{friedman1948utility}; risk-sensitive foragers switch between
	risk aversion and risk seeking as the foraging environment and their
	own energetic state vary \cite{kacelnik1996risky}. The mechanism
	behind is usually explained by the expected utility theory
	(EUT)\footnote{In EUT, utility measures the pleasure one enjoys from
	the money or resource received, which means utility is defined as a
	function of the rewards, denoted by $U(x)$, where $x$ represents the
	size of the reward. It is generally believed that most people are
	risk-averse in their daily life, for them $U(x)$ is a monotonically
	increasing convex function of $x$; and it is common to assume that
	the utility function of risk averters takes the form of logarithmic
	function. But the exact form of an individual's utility function
	varies from person to person and from case to case. For example,
	\$100 means nothing to a billionaire, but a poor man may associate
	it with rather high utility. For centuries, economists believe that
	when facing with a choice among portfolios with uncertain results
	(or portfolios containing risky assets), a rational person should
	choose the one that has the highest expected utility to him, rather
	than the one with the highest expected monetary value, that is, if
	$x$ is a random variable, then a personal person should pursue the
	maximization of $\left< U(x) \right>$ (where the symbol $\left<
	\cdot \right>$ denotes the mathematical expectation), rather than
	the maximization of $\left <x \right>$. This is the core idea of
	EUT.} \cite{neumann1944theory, savage1954foundations}, which is
	generally considered the orthodox theory of decision-making under
	uncertainty.
	
	Although EUT seems to be reasonable, in the past few decades, it has
	been severely challenged because of its failures in explaining some
	major theoretical issues \cite{ellsberg1961risk,rabin2001anomalies},
	among which the failure of EUT in justifying the anomalous change of risk attitude towards different levels of risk \cite{rabin2000risk} (mostly represented by the Allais paradox \cite{allais1953,slovic1974who}) have become the focus of attention. As a result, a series of non-expected utility
	theories have been developed mainly based on the assumptions of
	subjective weighted probabilities \cite{karmarkar1978subjectively}
	and decision heuristics \cite{starmer2000developments}, including
	prospect theory \cite{kahneman1979prospect,tversky1992advances},
	rank dependence \cite{quiggin1982theory}, bounded
	rationality\cite{simon1957}, salience theory
	\cite{bordalo2012salience}, etc.
	
	The discussion of risk attitude has also intrigued some geneticists,
	who have found that genetic influence may play a role in individuals'
	risk preference
	\cite{kreek2005genetic,zhong2009heritability,linner2019genome},
	showing that patterns of risk preference we see in both humans and
	animals are very likely the result of evolution, just like many
	other behaviors. Such findings
	provide genetic foundation for the engagement of evolution in the
	study of risk preference. Risk-sensitive foraging theories suggest
	that individuals switch between risk aversion and risk proneness for
	the purpose of maximizing their fitness, which is usually modeled as
	a nonlinear function of an individual's energy state, their
	capacity of energy reserve and the predation risk in the
	environment\cite{stephens1987,mcnamara1992risk}. More direct
	biological explanations of the utility function have been developed
	by considering the optimal solution in a game of evolution in which
	the payoff is determined by repeated
	lotteries\cite{cooper1982adaptive,karni1986self,yoshimura1991individual,robson1996biological,netzer2009evolution,brennan2011the,robson2001the,yoshimura2013dynamic1,zhang2014origin,levy2015an,robatto2017on}
	or in a competition for status\cite{dekel1999on,robson2001the},
	The mechanism of dynamics is extensively used in many of those studies, which track the change of behavior frequencies caused by both genetic heritage and imitation among individuals. For
	example, Zhang \emph{et al.}\cite{zhang2014origin} found through
	evolutionary dynamics that randomizing between risk-aversion and
	risk-neutrality can be optimal in the presence of both environmental
	uncertainty and idiosyncratic risk.
	
	In this paper, under the framework of evolutionary dynamics, we
	proposed the theoretical model of evolutionary rationality (ER), showing
	that it is the intensity of selection of an risky event that determines individuals' risk attitude towards it. An option is called an ER option for an individual if its geometric mean of fitness for that individual is at least not lower than that of all other possible options. Moreover, for the humans and economic activities, we believe that it is the mental equivalent of the intensity of selection, the attention degree one gives to a risky event, holds the key of her preference over different options, and the one with the highest (or equally highest) geometric mean of "fitness" for her is the ER option. As an example, we apply the concept of ER to the Allais paradox to justify the "anomaly" that is against the prediction of EUT.

	\section{The model and analysis}
	
	We consider an infinite population with discrete, nonoverlapping
	generations. Individuals in the population produce asexually, and
	their fitness (or number of offsprings) is determined by the choice they make only once in their life,
	which is to choose between two lotteries, A and B, with a reward of
	$a_t$ and $b_t$ respectively. To be more precise, the fitness of an
	individual that chooses A at any time step $t \ge 0$ is defined as
	\begin{eqnarray}
	\pi_{A,t} &=& W(1-w) +w a_t \ ,
	\end{eqnarray}
	and, similarly, the fitness of an individual who chooses $B$ is
	\begin{eqnarray}
	\pi_{B,t} &=& W(1-w) +w b_t \ .
	\end{eqnarray}
	
	Here $W$ represents the initial endowment inherited by an individual. Because we aim to find the best strategy for a given individual, we assume homogenous population that consists of "clones" of the same individual, who only differ in the choice they make. Therefore for simplicity we assume $W=1$ for everyone before they make their choices. The parameters $a_t$ and $b_t$, as rewards for lottery A and B, are assumed to be two non-negative random variables whose probability distributions are known to all individuals. $a_t$ and $b_t$ are not necessarily independent of each other. The mean and variance of $a_t$ are denoted by $\left<a_t\right>=\bar{a}$ and $\left< (a_s-\bar{a}) (a_t-\bar{a})\right>=\sigma_{a}^2\delta(t-s)$; similarly, the mean and variances of $b_t$ are $\left<b_t \right>=\bar{b}$ and $\left < (b_s-\bar{b}) (b_t-\bar{b}) \right> \sigma_b^2 \delta(t-s)$. The parameter $w \in (0,1)$ , under the framework of evolutionary dynamics, denotes the intensity of selection for an individual\cite{nowak2004emergence}. Intensity of selection measures the contribution of a reward to the total fitness of an individual, hence it can be regarded as the measure of importance
of an event or a behavior (with its consequence) to the individual. When $w$ approaches 1, it means the choice faced by the individual is a matter of life and death. When $w$ is close to 0, then the choice can barely affect the individual's total fitness. In the following paragraphs we will prove that $w$ holds the key of the transition of choices made in different cases and by individuals in different situations, and show that it establishes a fundamental connection between risk aversion (corresponding to large $w$) and risk neutrality (corresponding to small $w$).

	Let $x_t$ be the frequency of $A$-individuals (i.e., the proportion of
	individuals choosing $A$ in the population) at time step $t \ge 0$.
	Following the standard model of population evolutionary biology, the
	frequency of $A$-individuals at time step $t+1$, denoted by $x_{t+1}$, can be
	given by a stochastic recurrence equation
	\begin{eqnarray}
	x_{t+1} &=& \frac{x_t \pi_{A,t}}{x_t \pi_{A,t} +\big (1-x_t\big)
		\pi_{B,t}} \ = \ \frac{x_t \big ((1-w) +w a_t \big)}{(1-w) +w \big
		(x_t a_t +(1-x_t) b_t \big )}
	\end{eqnarray}
	with boundaries $\hat{x}=0$ and $\hat{x}=1$ (also called the
	fixation states of the system).  This equation describes the time evolution of the frequencies of $A$- and $B$-individuals based on $\pi_{A,t}$ and $\pi_{B,t}$. In fact, Eq. (3) can also be equivalently expressed as
	\begin{align}
	\Delta x_t &= x_{t+1}-x_t = x_t (1-x_t) \frac{\pi_{A,t}-\pi_{B,t}}{x_t \pi_{A,t} +(1-x_t) \pi_{B,t}} \ .
	\end{align}
	This implies that the evolutionary direction of $x_t$ is independent of the system state at any time step $t$, that is, the time evolution of $x_t$ is assumed to be frequency-independent in our model. For our main goal, we are interested in the stochastic stability of the fixation states $\hat{x}=0$ and $\hat{x}=1$.
	
	Following Karlin and Liberman \cite{karlin1974random} (see also\cite{zheng2017evolutionary}), a fixation state $\hat{x}$ of Eq. (3) is said to be
	stochastically locally stable (SLS) if for every $\epsilon >0$ there
	exists $\delta_0 >0$ such that $\mathbb{P} (x_t \rightarrow \hat{x})
	\ge 1-\epsilon$ as soon as $|x_0-\hat{x}| < \delta_0$. On the other
	hand, a fixation state $\hat{x}$ is said to be stochastically
	locally unstable (SLU) if $\mathbb{P}(x_t \rightarrow \hat{x})=0$ as
	soon as $|x_0-\hat{x}|>0$. Note that the time evolution of $x_t$ is frequency-independent. Thus, if the fixation state $\hat{x}=1$ (or $\hat{x}=0$) is SLS, then, for any possible initial value $x_0$ being in the interval $0<x_0<1$, $A$-individuals (or $B$-individuals) will eventually occupy the whole population with a probability close to $1$. 
	
	Defining $u_t=x_t \big / \big (1-x_t \big)$, Eq. (3) can take a
	simple form
	\begin{align}
	u_{t+1} &= u_t \ \frac{(1-w) +w a_t}{(1-w) +w b_t} \ .
	\end{align}
	We here present only the simplified mathematical arguments for the
	stochastic local stability of $\hat{x}=0$, which corresponds to
	$\hat{u}=\hat{x} \big / (1-\hat{x}) =0$ (the more rigorous
	mathematical proofs are similar to those in \cite{zheng2017evolutionary}.
	Iterating the stochastic recurrence equation in (5) leads to
	\begin{align}
	& u_n = u_0 \prod_{t=0}^{n-1} \left [ \frac{(1-w)+w a_t}{(1-w) +w b_t} \right] \nonumber \\
	\Rightarrow & \ \frac{1}{n} \Big [ \log u_n - \log u_0 \Big ] =
	\frac{1}{n} \sum_{t=0}^{n-1} \log \left [ \frac{(1-w)+w a_t}{(1-w)
		+w b_t} \right]
	\end{align}
	for $n \ge 1$. For large $n$, the strong law of large numbers
	guarantees that
	\begin{align}
	\lim_{n \rightarrow \infty} \frac{1}{n} \Big [ \log u_n -\log u_0
	\Big] &= \left<\log \left [\frac{(1-w)+w a_t}{(1-w) +w b_t} \right ]
	\right> \ .
	\end{align}
	Using Egorov's theorem, it can be shown that $\hat{x}=0$ is SLS if
	\begin{eqnarray}
	\left<\log \left [\frac{(1-w)+w a_t}{(1-w) +w b_t} \right ] \right>
	&<& 0 \ .
	\end{eqnarray}
	This means that $\hat{x}=0$ is SLS if the geometric mean of $\pi_{A,t}$ is less than that of $\pi_{B,t}$, that is, $\left< \log \pi_{A,t} \right> < \left< \log \pi_{B,t} \right>$. Similarly, the fixation state $\hat{x}=1$ is SLS if
	\begin{eqnarray}
	\left<\log \left [\frac{(1-w)+w a_t}{(1-w) +w b_t} \right ] \right>
	&>& 0 \ ,
	\end{eqnarray}
	that is, $\left< \log \pi_{A,t} \right> > \left< \log \pi_{B,t} \right>$. 
	
	Based on the above analysis of stochastic evolutionary dynamics, we have the following definition:
	
	\textbf{\emph{Definition}.} \emph{The option $A$ (or $B$) is said to be an ER option if the geometric mean of $\pi_{A,t}$ (or $\pi_{B,t}$) is larger than that of $\pi_{B,t}$ (or $\pi_{A,t}$). As a special case, if the geometric means of $\pi_{A,t}$ and $\pi_{B,t}$ are the same, i.e. $\left< \log \pi_{A,t} \right> = \left< \log \pi_{B,t} \right>$, then options $A$ and $B$ should both be ER options.}
	
	For the situation with that both the variances of $a_t$ and $b_t$, $\sigma_a^2$ and $\sigma_b^2$, are not large, we have the approximations
	\begin{align}
	\left< \log \pi_{A,t} \right> &\approx \log \big ((1-w) +w \bar{a} \big) -\frac{w^2 \sigma_a^2}{2 \big ((1-w) +w \bar{a} \big )^2} + \cdots \ , \nonumber \\
	\left< \log \pi_{B,t} \right> &\approx \log \big ((1-w) +w \bar{b} \big) -\frac{w^2 \sigma_b^2}{2 \big ((1-w) +w \bar{b} \big )^2} + \cdots \ .
	\end{align}
	Therefore, option $A$ is an ER option if
	\begin{align}
	\log \left [ \frac{(1-w) +w \bar{a}}{(1-w) +w \bar{b}} \right ] &> \frac{w^2}{2} \left [ \frac{\sigma_a^2}{\big((1-w) +w \bar{a} \big)^2} - \frac{\sigma_b^2}{\big((1-w) +w \bar{b} \big)^2} \right ] \ ;
	\end{align}
	and, similarly, option $B$ is an ER option if
	\begin{align}
	\log \left [ \frac{(1-w) +w \bar{b}}{(1-w) +w \bar{a}} \right ] &> \frac{w^2}{2} \left [ \frac{\sigma_b^2}{\big((1-w) +w \bar{b} \big)^2} - \frac{\sigma_a^2}{\big((1-w) +w \bar{a} \big)^2} \right ] \ .
	\end{align}
	
	It is obvious that $w$ plays a key role in the criteria of the ER option: (\emph{\textbf{i}}) when $w$ is near to $1$ but $w \ne 1$, $\left<\log \pi_{A,t} \right> \approx \left< \log a_t \right>$ and $\left<\log \pi_{B,t} \right> \approx \left< \log b_t \right>$; and (\emph{\textbf{ii}}) when $w$ is near to $0$ but $w \ne 0$, $\left<\log \pi_{A,t} \right> \approx w \bar{a}$ and $\left<\log \pi_{B,t} \right> \approx w \bar{b}$. Therefore, if $w$ is large, then the option $A$ (or $B$) is an ER option if $\left< \log a_t \right> >\left<\log b_t \right>$ (or $\left< \log a_t \right> < \left<\log b_t \right>$); and if $w$ is small, then the option $A$ (or $B$) is an ER option if $\bar{a} >\bar{b}$ (or $\bar{a}< \bar{b}$). Specifically, if both $a_t$ and $b_t$ are discrete random variables, which are defined as $a_t=a_i$ with probability $p_i$ for $i=1,2, \cdots,n$ ($\sum_{i=1}^n p_i=1$) and $b_t=b_j$ with probability $q_j$ for $j=1,2,\cdots,m$ ($\sum_{j=1}^m q_j=1$), respectively, at any time step $t \ge 0$, then
	\begin{itemize}
		\item[(\textbf{a})]for $w \rightarrow 1$, the condition $\left<\log a_t \right> > \left< \log b_t \right >$ (or $\left<\log a_t \right> < \left< \log b_t \right >$) is equivalent to $\prod_{i=1}^n a_i^{p_i} > \prod_{j=1}^m b_j^{q_j}$ (or $\prod_{i=1}^n a_i^{p_i} < \prod_{j=1}^m b_j^{q_j}$);
		\item[(\textbf{b})]for $w \rightarrow 0$, the condition $\bar{a} > \bar{b}$ (or $\bar{a}<\bar{b}$) is equivalent to $\sum_{i=1}^n p_i a_i > \sum_{j=1}^m q_j b_j$ (or $\sum_{i=1}^n p_i a_i < \sum_{j=1}^m q_j b_j$).
	\end{itemize}
	
	\textbf{\emph{Evolutionary rationality in the human world.}} It has
	been well established in behavior genetics
	\cite{bouchard2003genetic} and evolutionary psychology
	\cite{buss2019} that many human behaviors and their mental basis are
	shaped by selection. In the human society, although fitness is no
	longer our first concern, it is very likely that we somewhat still
	behave the same way as if our fitness is at stake, because the
	preference we have for risk is deeply affected by the genetic
	heritage we received from our ancestors \cite{cesarini2009genetic}.
	
	Under such assumption, we call the projection of fitness in our mind
	the "meta-fitness". And the intensity of selection $w$, which might be
	gradually internalized and mentally processed ever since the
	beginning of human civilization, becomes the attention degree that
	one gives to a certain risky event. As a result, a person who is
	evolutionary rational will try to maximize the geometric mean of the
	meta-fitness generated by her choics. Specifically, if an
	individual associate an event with a rather small $w$, it probably means that
	the possible consequences (or the difference of the consequences) of the event are not important to her, or
	the possibility that something important is going to happen is
	extremely low. In such a case, the evolutionary rational individuals will incline to risk
	neutrality and pursue the highest arithmetic mean among the
	available choices. Contrarily, a high $w$ means that the individual
	think the impact of the event is profound; it is only natural that
	they proceed with discretion and be more averse to risk by choosing
	the option with the highest geometric mean. Obviously, $w$ varies
	for different cases and different individuals. A bet of \$ 1 and \$
	1 million have different importance to everyone; on the other hand,
	a billionaire will not consider a bet of \$ 1,000 the same way as
	an average person does. In the next section we will show the
	application of evolutionary rationality in economic activities with
	the examples of Allais paradox, which demonstrates the
	transition between extreme risk aversion and risk neutrality.

	\textbf{\emph{Evolutionary rationality of Allais Paradox.}} The
	Allais paradox proposed in 1953 represents one of the most
	well-known violation of EUT \cite{allais1953}. It arises from a
	binary-choice model containing two scenarios. Imagine you are
	offered to choose from the options in Table 1. For those who are
	risk-seeking or risk-neutral, the decision is obvious. But for the
	risk-averse majority, the paradox appears. Economists believe that
	most people prefer $A$ than $B$ in scenario 1, and $D$ than $C$ in scenario
	2. However, EUT predicts that one should either prefer both $A$ \& $C$ ,
	or both $B$ \& $D$, because EUT assumes linearity in
	probability weighting, and the options $C$ than $D$ in scenario 2 are derived
	directly from those of scenario 1 by deducting the 89\% probability
	of 100 million gain from $A$ and $B$ respectively. Therefore, the
	contradiction of theoretical prediction of EUT and empirical results
	created the Allais paradox.
	
	\begin{table}[h]
		\centering
		\caption{Allais paradox: the original version}
		\begin{tabular}{lllllll}
			\hline \hline
			\multicolumn{3}{l}{SCENARIO 1: Choose between}           &       &                        &                              &       \\ \hline
			& \multicolumn{1}{r}{\emph{A}:} & \$100 million with certainty   &       & \multicolumn{1}{r}{\emph{B}:} & \$500 million with probability & 0.10, \\
			&                        &                              &       &                        & \$100 million with probability & 0.89, \\
			&                        &                              &       &                        & \$0 with probability         & 0.01; \\
			&                        &                              &       &                        &                              &       \\ \hline \hline
			\multicolumn{3}{l}{SCENARIO 2: Choose between}           &       &                        &                              &       \\ \hline
			& \multicolumn{1}{r}{\emph{C}:} & \$100 million with probability & 0.11, & \multicolumn{1}{r}{\emph{D}:} & \$500 million with probability & 0.10, \\
			&                        & \$0 with probability         & 0.89; &                        & \$0 with probability         & 0.90. \\
			&                        &                              &       &                        &                              &       \\ \hline
		\end{tabular}
	\end{table}
	
	It is obvious that the original Allais paradox is too unrealistic.
	To further verify the existence of Allais paradox, Kahneman and Tversky conducted a questionnaire survey on the
	paradox with much lower prize \cite{kahneman1979prospect}, which is
	shown in Table 2. Their survey showed that 82\% subjects chose $A$
	over $B$ in scenario 1, and 83\% $D$ over $C$ in scenario 2, proving that the paradox did
	exist.Similar behavioral patterns have also been found in animal
	experiments\cite{battalio1985animals,real1996paradox}.
	
	When we put the KT version of Allais paradox under the framework of
	evolutionary rationality, we can see that, in scenario 1, option
	$A$ is said to be evolutionarily rational if
	\begin{eqnarray}
	\log \big (W(1-w) +2400w \big) &>& 0.33 \times \log \big (W(1-w)+2500w \big) \nonumber \\
	&& + 0.66 \times \log \big (W(1-w) +2400w \big) \nonumber \\
	&& +0.01 \times \log(W(1-w)) \ ;
	\end{eqnarray}
	and in scenarior 2, option $D$ is said to be evolutionarily
	rational if
	\begin{eqnarray}
	&& 0.33 \times \log \big ( W(1-w) +2500w \big ) +0.67 \times \log(W(1-w)) \nonumber \\
	&>& 0.34 \times \log \big ( W(1-w) +2400w \big ) + 0.66 \times
	\log(W(1-w)) \ .
	\end{eqnarray}
	
	\begin{table}[h]
	\centering
	\caption{Allais paradox: the KT version}
	\begin{tabular}{lllllll}
		\hline \hline
		\multicolumn{3}{l}{SCENARIO 1: Choose between}           &       &                        &                              &       \\ \hline
		& \multicolumn{1}{r}{\emph{A}:} & \$2400 with certainty   &       & \multicolumn{1}{r}{\emph{B}:} & \$2500 with probability & 0.33, \\
		&                        &                              &       &                        & \$2400 with probability & 0.66, \\
		&                        &                              &       &                        & \$0 with probability         & 0.01; \\
		&                        &                              &       &                        &                              &       \\ \hline \hline
		\multicolumn{3}{l}{SCENARIO 2: Choose between}           &       &                        &                              &       \\ \hline
		& \multicolumn{1}{r}{\emph{C}:} & \$2400 with probability & 0.34, & \multicolumn{1}{r}{\emph{D}:} & \$2500 with probability & 0.33, \\
		&                        & \$0 with probability         & 0.66; &                        & \$0 with probability         & 0.67. \\
		&                        &                              &       &                        &                              &       \\ \hline
	\end{tabular}
\end{table}
	
	The solution of the inequality in (13) and (14) is shown in Fig. 1. The blue curve represents the $w^*$ that equalize the two sides of (13) and (14) for any given $W$. Above the curve, the blue area covers the solution of (13); and below the curve, the orange area shows the solution of (14). For example, for an individual with an endowment that is equivalent to \$14,000 (represented by the red vertical line), she will choose both option $A$ and $D$ if in scenario 1 she has a $w>w^*=0.843$, and in scenario 2 a $w<w^*=0.843$. More generally, the survey results of Kahneman and Tversky suggest that: (\textbf{\emph{i}}) in scenario 1, most people allocate heavy attention on the lottery (i.e. $w>w^*$), probably because a guarantee of of \$2400 is important income for them, while choosing $B$ risks getting nothing in return; and (\textbf{\emph{ii}}) in scenario 2, most people does not pay extremely high attention (i.e. $w<w*$) to the lottery. In addition, we present the solution of the original Allais paradox in Fig. 2. The result shows that for an evolutionary rational individual who chooses option $A$ in scenario 1, the attention she pays to the lottery must be extremely high. And in scenario 2, as long as the attention degree given to the lottery by a person is not close to 1, option $D$ will become the ER option for her. Therefore, the Allais paradox can be well explained by the evolutionary rationality theory of risk preference.
	
	\includegraphics{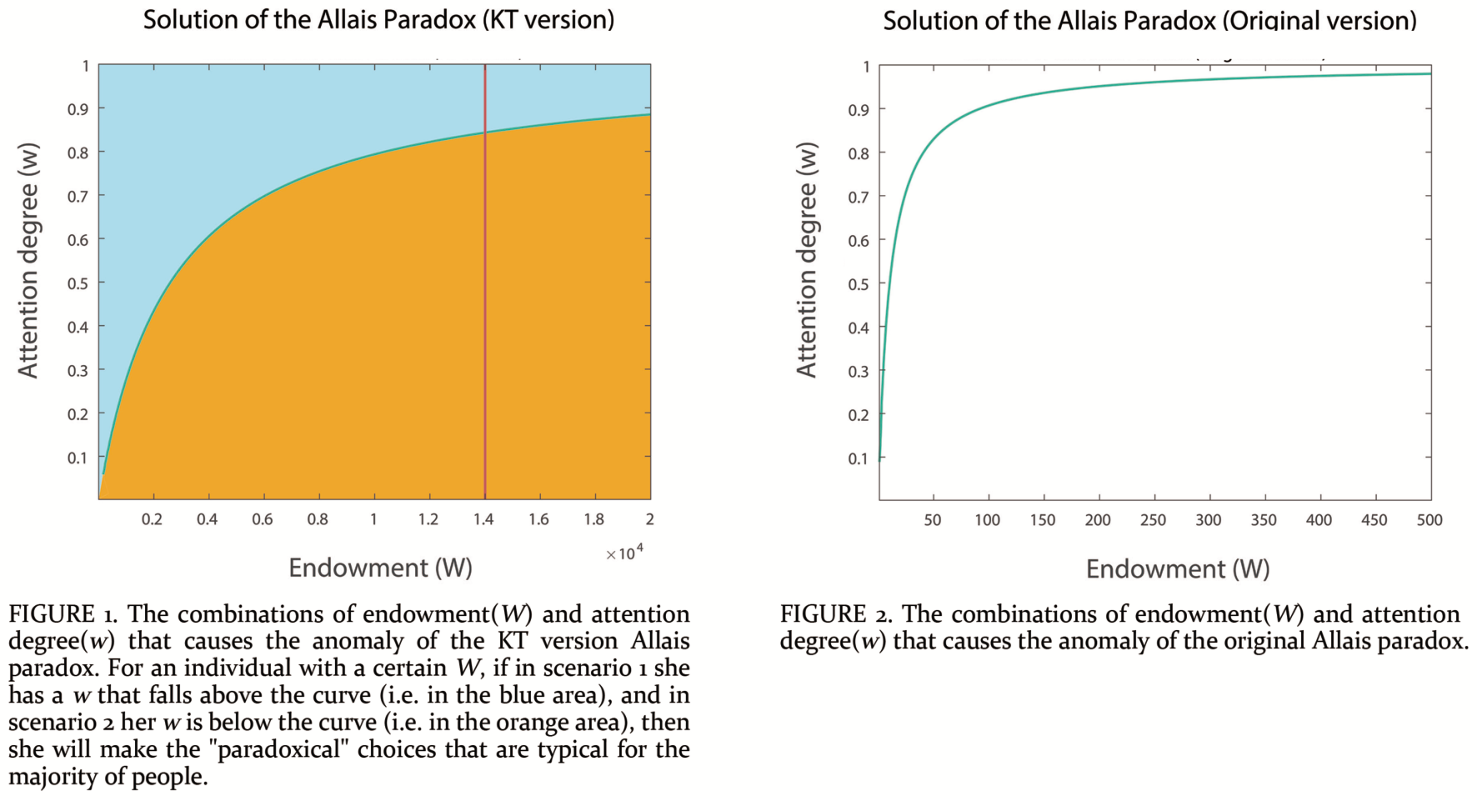}

	\section{Discussion}
	
	Since the solution of St. Petersburg paradox was first proposed in the 18th century, researchers have found that decision making under uncertainty constitutes an important part of an individual's behavior pattern, and many theories have been developed in the fields of economics and psychology. In recent decades, risk-sensitive behaviors have been recognized in many animal species, showing that the same patterns of risk preference we have in the human society are also shared by animals with limited cognitive ability. Therefore, we believe the risk preference that prevails in the majority of individuals (including both humans and animals) must have been successful in evolution, for which it should be flexible yet optimal under different pressure of selection. In this paper, we
	engaged stochastic population evolutionary dynamics to find the risk
	preference that was advantageous in evolution, and proposed the
	theory of evolutionary rationality. Intensity of selection (measured by
	the parameter $w$) was introduced as a measure of importance of a
	risky event to an individual. With the help of it, the fitness of an
	individual using a certain strategy was determined, and the
	stability analysis of evolutionary dynamic analysis found that the
	strategy of always choosing the option with the highest geometric
	mean of fitness would spread in the population, and became the final
	winner in evolution.
	
	Then it follows the question of how we, the human beings who no
	longer struggle for survival and reproduction, perceive selection
	intensity and fitness in our mind, especially when we are making
	decisions for economic activities. First, the intensity of selection,
	we believe, can be understood as the attention degree that
	transforms the final payoff of an event to its true mental value
	evaluated in the process of decision making. Then the fitness,
	although not supported by any solid evidence, was interpreted in
	many researches as utility \cite{gandolfi2018}, the satisfaction one
	feels for the potential gain. Such an interpretation certainly
	sounds plausible, because if utility is truly behind every economic
	decision we make, there ought to be an evolutionary basis for it.
	Nonetheless, we are uncertain whether it is true for our theory due
	to the lack of supportive arguments of any kind. Therefore we simply
	call it the "meta-fitness" and define it as the sum of a person's
	endowment before the decision making and the payoff they receive
	from the risky event weighted by the attention degree they give to
	it. Similarly, the maximization of the geometric mean of the
	meta-fitness will become the pursuit of a person who is evolutionary
	rational. And the heuristic and simplified rule for such pursuit can
	be made without involving fitness or meta-fitness: adopt extreme
	risk aversion and choose the option with the highest geometric mean
	of the payoffs if you think the event is very important, and adopt
	risk neutrality and choose the option with the highest arithmetic
	mean if you think the possibilities of the event and their
	corresponding results barely matter.
	
	We have noted that the allocation of attention has always been an
	important aspect of many economic theories about risk attitude,
	especially those trying to explain the "irrational behaviors" that
	are adopted by the majority of people, among which the Allais
	paradox, as the typical example, has received the most attention in
	those researches. The prospect
	theory\cite{kahneman1979prospect,tversky1992advances} believes that
	excessive attention is given to low probabilities so that people
	tend to overweight the occurrence of rare consequence, and
	underweight the probability of common events. On the other hand,
	salience theory\cite{bordalo2012salience} argues that attention is
	biased among outcomes. The outcome that differs most from the other
	ones attracts the most attention, highlighting either the upside or
	the downside of the lottery in one's mind, and therefore determines
	an individual's risk preference in it. In our model, attention
	degree is derived from a more fundamental concept, the selection
	intensity, which, works on an event and the choice making behavior
	in response to that event as a whole. In the rather long journey of
	selection, every possible result of a choice is going to happen at
	some point, and selection will work fairly on every one of them,
	without bias towards particular possibilities or outcomes. It is
	also worth noting that because the decision of choice that is under
	selection is made upon comparison of fitness or meta-fitness, not on
	the exact value of them, we consider fitness and meta-fitness
	ordinal but not cardinal, which is similar to what is widely
	believed of the utility.
	
	Although risk preference is usually considered an economic problem,
	evolutionary dynamics has contributed with great insights to this
	area. A series of theoretical works have been developed to justify
	the existence of utility function, which is interpreted as the
	functions of fitness or survival chance of an individual or a
	lineage. Naturally many components of the evolutionary process have
	also been found crucial in the shaping of the utility function. The
	most noteworthy examples are the risk
	types\cite{yoshimura1991individual,robson1996biological,brennan2011the,zhang2014origin,robatto2017on}
	and mutation\cite{robson1996biological,brennan2012variety}. However, few efforts have been made to explain the Allais paradox, even though the same anomalous behaviors have also been found in animal experiments\cite{battalio1985animals,real1996paradox}. We are only aware of one that works to accommodate the anomaly of Allais paradox, in which the framing effect following a shift of the reference point is provided as the main explanation for the anomaly\cite{yoshimura2013dynamic2}. There is no doubt that framing effect is a prominent cognitive bias with deep implications in economic behaviors, but it is not required in our model to explain Allais paradox. With the help of a simple parameter $w$, we are able to show the evolutionary rationality of the behavior pattern shared by most people. As different risky events attract different attention degree, an individual's behavior is bound to change between extreme risk aversion to risk neutrality.

	\newpage
	
	\bibliographystyle{unsrt}

\end{document}